\documentclass[10pt,preprint]{aastex}
\def\smtopskip{-1.2cm}
\def\smbotskip{-1.0cm}
\usepackage{emulateapj5}
\usepackage{danonecolfloat}
\usepackage{epsfig}

\newcommand{\avg}[1]{\ensuremath{\langle{#1}\rangle}}

\begin{document}

\def\nyu{1}

\twocolumn[%%% Begin front material
\title{A simple model for the clustering of subhalos as a function of
       mass}
\author{
  Michael~R.~Blanton\altaffilmark{\nyu}
}

\begin{abstract}
Astrophysicists do not associate galaxies with the massive dark matter
halos so easily detected in particle simulations and whose mass
function can be predicted by the Press-Schechter formalism and its
variants. Instead, they associate galaxies with smaller surviving
subhalos whose contents do not fully mix into the larger dark matter
halos they exist in. However, these subhalos are as of now difficult
to find in simulations.  I present a simple description of what the
Press-Schechter formalism can say about the clustering of these
subhalos, under the restrictive assumption that at fixed subhalo mass,
the chance of survival of the subhalo is independent of its larger
scale environment. This model predicts a weak dependence of clustering
on mass below $M \approx 10^{12} M_\odot$, in agreement with
observations of a weak dependence of clustering on luminosity at
luminosities less than $L_\ast$. The model is in qualitative agreement
with observations that at low luminosities galaxy clustering is a
strong function of color while at high luminosities galaxy clustering
is a strong function of luminosity.
\end{abstract}

]%%% End front material

\section{Motivation}

\footnotetext[\nyu]{Center for Cosmology and Particle Physics, Department of Physics, New
York University, 4 Washington Place, New York, NY 10003
\texttt{blanton@nyu.edu}}

The excursion set formalism of \citet{press74a} (and the subsequent
improvements of \citealt{bond91a}, \citealt{lacey93a}, and
\citealt{sheth01a}) predicts the mass function of bound, virialized
objects in the universe. The extensions of \citet{mo96a} and
\citet{sheth01a} predict the clustering relative to dark matter of
these halos. However, the large-scale, sufficiently high resolution
$N$-body calculations of \citet{kravtsov99a} and other investigators
show that each CDM dark matter halo can contain multiple subhalos,
presumably ex-halos not fully mixed into the larger, surrounding
halo. The halos themselves do not have properties similar to those of
galaxies --- most notably, they grow to be much more massive. However,
it is plausible that galaxies reside in subhalos.

Here I predict the clustering of subhalos as a function of redshift,
mass, and scale, using a modification of the method of
\citet{mo96a}. I use the formation rate of halos to determine when to
identify halos of a particular mass and then evaluate the clustering
of those halos at the redshift of observation.  Where cosmological
parameters are necessary for these calculations I use those given by
\citet{bennett03a} in their Table 3.

\section{Estimating the clustering of subhalos of a specific mass}

Consider a set of subhalos which when they reach mass $M$ are
identifiable galaxies that retain their unique identities throughout
history until $z=0$, even after they merge into larger halos. Let us
make one simplifying assumption: whether a subhalo of a given mass
retains its identity during mergers into larger halos is independent
of when it formed and its large scale environment. In other words, of
all subhalos of mass $M$ which ever existed, those which survive form
a representative sample.  This assumption is violated in the real
universe and in simulations, but it provides an alternative to the
standard treatment, in which each subhalo always loses it identity
immediately after its formation.

One can calculate the dependence on redshift of the formation rate of
such subhalos from the extended Press-Schechter theory, the clustering
at each redshift using the {\it ansatz} of \citet{mo96a}, and
integrate the results to the observed redshift $z_{\mathrm{obs}}$
using the continuity equation
(\citealt{nusser94a,fry96a,tegmark98a}). Notably, the same procedure
cannot predict the abundance of subhalos as a function of mass,
without further specification of the conditions of subhalo formation
and survival (\citealt{sheth97a,benson01a,somerville01a}).

The Press-Schechter formulae are derived by assuming that the
primordial overdensity field $\delta$ is a Gaussian random field.  The
smoothed density field is distributed as a Gaussian with variance
$\sigma^2(M)$, where $M$ corresponds to the average mass contained
within a smoothing length $R$. One associates each parcel of mass in
the original distribution to halos of mass $M$ at a particular
redshift $z$ {\it if} at that point in space, the smoothed density
field on scale $R$ linearly extrapolated (using the growth function
$D(z)$) to $z$ is exactly some $\delta_{c,0}$. The conventional choice
is $\delta_{c,0} = 1.7$. We define $\delta_c(z) = \delta_{c,0}/D(z)$.
The fraction of mass in such a state is
\begin{eqnarray}
\label{fden}
f(M,z) dM 
&=& \frac{1}{\sqrt{2\pi}} \frac{\delta_c(z)}{\sigma^3(M)} \left[-
\frac{d\sigma^2(M)}{dM} \right] \cr
&& \quad\times \exp\left[- \delta_c^2(z)/2\sigma^2(M)
\right] dM .
\end{eqnarray}

\cite{lacey93a} derive expressions for the probability of merging
between dark matter halos.  The fraction of mass in halos of mass
$M_2$ which just entered from halos of mass $M_1$ is:
\begin{eqnarray}
\frac{d^2f(M_1|M_2, z)}{dz dM_1} &=& \frac{1}{\delta_c}
\frac{1}{\left[1-\sigma^2(M_2)/\sigma^2(M_1) \right]^{3/2}} \cr
&&
\times \exp\left[\frac{\delta_c^2}{2\sigma(M_1)}
\right]
\left(-\frac{d\delta_c}{dz}\right) f(M_1,z).
\end{eqnarray}

The formation rate of mass $M$ halos is proportional to the fraction
of mass in $M$ halos which has entered recently from all smaller
halos. One can calculate this from the fraction entering $M$ from each
smaller mass halo, integrated over all possible smaller mass halos:
\begin{eqnarray}
\label{fdenform}
\frac{df_\mathrm{form}(M,z)}{dz} 
&=& \int_0^M dM_1 \frac{df(M_1|M, t)}{dzdM_1} f(M)\cr
%&=& \frac{1}{\sqrt{2\pi}} 
%\left(-\frac{d\delta_c}{dt}\right) f(M,t) \cr
%&& \quad \times \int_{\sigma^2(M)}^{\infty} d(\sigma^2(M_1)) 
%\frac{1}{\left[\sigma^2(M_1) - \sigma^2(M)
%\right]^{3/2}}\cr
&=& \frac{1}{\sqrt{2\pi}} 
\left(-\frac{d\delta_c}{dz}\right) f(M,z) \cr
&& \times \lim_{M_1\rightarrow M}
\frac{2}{\left[\sigma^2(M_1) - \sigma^2(M)
\right]^{1/2}}
\end{eqnarray}
The divergence is irrelevant to the final result; below, I will
discuss only the ``shape'' of the formation rate, which I express as
\begin{equation}
\frac{dG}{dz} = \delta_c(z)
\left(-\frac{d\delta_c}{dz}\right)\exp\left(-\delta_c^2/[2\sigma^2(M)]\right).
\end{equation}

The Press-Schechter formalism has a simple prediction that the
clustering of halos of a particular mass is indendent of their
formation history. Thus the overdensity the newly born halos of mass
$M$ relative to their mean density is the same as the overdensity of
all halos of mass $M$ relative to their mean, for which \citet{mo96a}
obtain:
\begin{eqnarray}
\label{b_form}
1+\avg{\delta_h|\delta} &=&
\frac{1-\delta_0/\delta_c}{\left[ 1 -
\sigma^2(M_0)/\sigma^2(M)\right]^{3/2}}  \cr
&& \times \exp\left[-\frac{(\delta_c-\delta_0)^2}{2\left[\sigma^2(M) -
\sigma^2(M_0)\right]} + \frac{\delta_c^2}{2\sigma^2(M)}\right] \cr
&&\times \left\{ 1 -
\exp\left[\frac{\delta_0^2}{2\sigma^2(M_0)} 
- \frac{(\delta_0-2\delta_c)^2}{2\sigma^2(M_0)}
\right]
\right\}.
\end{eqnarray}

Assume a subhalo is in a mass overdensity $\delta$ at the redshift of
observation $z_{\mathrm{obs}}$, which determines the linearly
extrapolated overdensity $\delta_0$. From this, one can find
$\delta_{\mathrm{form}}(M,z_f,\delta_0) \equiv \avg{\delta_h|\delta}$
at formation for subhalos of mass $M$ formed at some redshift
$z_f>z_{\mathrm{obs}}$ which end up in mass overdensities of $\delta$
at redshift $z_{\mathrm{obs}}$. Then:
\begin{eqnarray}
\label{nlbias_int}
&& 1+\avg{\delta_{\mathrm{sh}} (M,z_{\mathrm{obs}})|\delta} = 
\frac{1+\delta(z_{\mathrm{obs}})}{G(M,z_{\mathrm{obs}})}\cr
& & \times \int_{z_{\mathrm{obs}}}^\infty dz_f
[1+\delta_{\mathrm{form}}(M,z_f,\delta_0) ]
\frac{dG(M,z_f)}{dz_f},
\end{eqnarray}
To calculate quantities in the nonlinear regime, one must relate
$\delta_0$ to $\delta$, which I do using the formulae given by
\citet{mo96a} for an Einstein-de Sitter universe (the solution to the
nonlinear spherical collapse problem is roughly independent of
cosmology).

The integrals above may be done explicitly to yield
\begin{eqnarray}
\label{nlbias}
&& 1+\avg{\delta_{\mathrm{sh}} (M,z_{\mathrm{obs}})|\delta} = \cr
&& \quad \frac{[1+\delta(z_{\mathrm{obs}})]
\exp\left(\delta_c^2/[2\sigma^2(M)]\right)}
{\sigma^2(M)
\left[1-\sigma^2(M_0)/\sigma^2(M)\right]^{3/2}}\cr
&& \quad \times \{ \Delta^2 
  \exp\left(-(\delta_c-\delta_0)^2/[2\Delta^2]
	\right) 
- \Sigma^2
	\exp\left(-x_c^2 + \Phi^2 \right) \cr
&& \quad\quad + \sqrt{2\pi} (\Sigma \Delta^2/\Upsilon^2) \delta_0
% \cr
%&& \quad \quad \times 
\exp\left(\Phi^2\right) \mathrm{erfc}(x_c) \}
\end{eqnarray}
where
\begin{eqnarray}
\Delta^2&=&\sigma^2(M)-\sigma^2(M_0) \cr
\Upsilon^2&=&4\sigma^2(M)-3\sigma^2(M_0) \cr
\Phi^2&=&2\delta_0^2 \Delta^2/[\sigma^2(M_0)\Upsilon^2]\cr
\Sigma^2&=&\sigma^2(M_0) [\Delta^2 /\Upsilon^2]\cr
x_c &=& \frac{\delta_c-(2\sigma^2(M)-\sigma^2(M_0))\delta_0/\Upsilon^2}
{\sqrt{2\Sigma^2}}
\end{eqnarray}

To calculate the cross-correlation \avg{\delta_{sh}\delta} I follow
\citet{mo96a} and use the log-normal form for the distribution of the
dark matter overdensity $\delta$:
\begin{eqnarray}
\label{avgbias}
	\avg{\delta_{sh} \delta} &=& \int_{-1}^\infty d\delta \delta \avg{\delta_{sh}|\delta}
	f(\delta) \cr
&=& \int_{-1}^\infty d\delta \frac{\exp(-x^2/[2\sigma_l^2])}
{\sqrt{2\pi} \sigma_l(1+\delta)}
\avg{\delta_{sh}|\delta} \delta,
\end{eqnarray}
where $x= \ln(1+\delta) + \sigma^2/2$ and
$\avg{\delta^2}=\exp(\sigma^2) -1$.

At linear scales the above equations can be reduced to a linear bias
of the form
\begin{equation}
\label{linbias}
b_{\mathrm{sh}}(z) = 1 + \frac{\delta_c(z)}{\sigma^2(M) D(z)},
\end{equation} 
which has the notable property that it is always greater than unity.

\def\figsize{13.5cm}
\def\figsiz{13.5cm}
\begin{figure} 
\vskip\smtopskip
\centerline{\epsfxsize=\figsize\epsffile{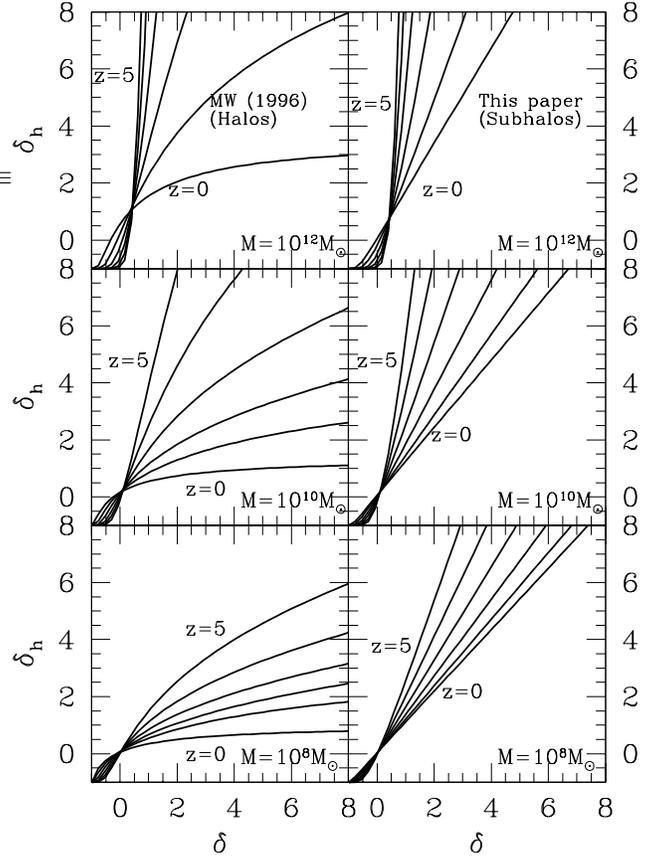}}
\vskip\smbotskip
\caption[1]{\label{delta}\footnotesize% 
The relationship between overdensity of
halos and overdensity of dark matter from the treatment of
\citet{mo96a} ({\it on the left}) and that from Equation
\ref{nlbias} for subhalos ({\it on the right}). This figure has
results for a smoothing scale $R = 4$ $h^{-1}$ Mpc, for masses $M =
10^8, 10^{10},$ and $10^{12} M_\odot$, as labeled, and for redshifts
$z=0,1,2,3,4,$ and $5$, as labeled. }
\end{figure}

\def\figsize{10.5cm}
\def\figsiz{10.5cm}
\begin{figure} 
\vskip\smtopskip
\centerline{\epsfxsize=\figsize\epsffile{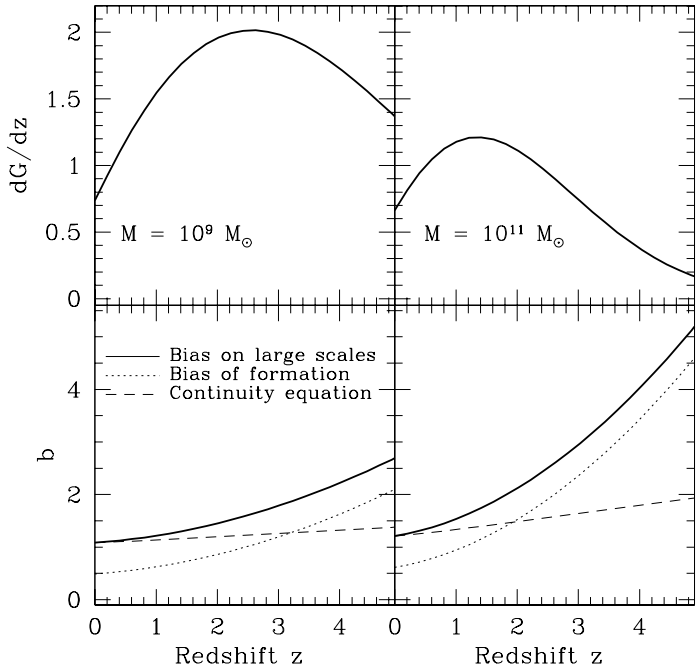}}
\vskip\smbotskip
\caption[1]{\label{biascomp}\footnotesize% 
{\it Top panels}: Rate of
formation $dG/dz$ of subhalos of mass $10^{9} M_\odot$ {\it (left)}
and $10^{11} M_\odot$ {\it (right)} in arbitrary units. {\it Bottom
panels}: The solid line is the linear bias on large scales based on
Equation \ref{linbias}. The dotted line is the ``bias at formation''
from Equation \ref{b_form}. The dashed line is how you would
extrapolate the results backwards from $z=0$ using only the continuity
equation. The result of Equation \ref{linbias} is not far from
assuming that all galaxies formed at near the median of the $dG/dz$
distribution and were merely debiased from then on. }
\end{figure}

\def\figsize{10.5cm}
\def\figsiz{10.5cm}
\begin{figure} 
\vskip\smtopskip
\centerline{\epsfxsize=\figsize\epsffile{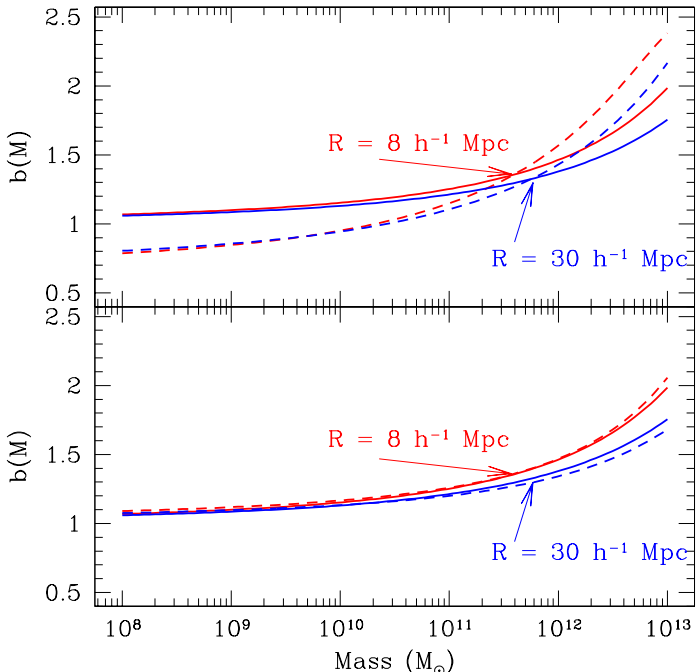}}
\vskip\smbotskip
\caption[1]{\footnotesize% 
Bias at redshift $z=0$ as a function of mass for three different
scales, as labeled.  The solid lines are the predictions of Equation
\ref{nlbias}. The dashed liens are the predictions from
\citet{mo96a} for halos identified at redshift $z=1.5$ and observed at
$z=0$.
\label{massdep}}

\end{figure}

Note that I have not specified what {\it overall number} of subhalos
end up at each mass! In fact, without further assumptions, one cannot
do so. I have only assumed that the subhalos that survive at each mass
have clustering properties representative of all the subhalos ever to
have that mass. I ignore that this condition is almost certainly
violated in the real universe.

\section{Results}

First consider Figure \ref{delta}, showing the halo overdensity as a
function of dark matter overdensity on scales of $R= 4$ $h^{-1}$ Mpc,
for three different halo masses, at redshifts $z=0$\ldots$5$, as
labeled. The left column is the results of \citet{mo96a} for halos
identified at the same redshift as they are observed. The right column
is the results of Equation \ref{nlbias}. Particularly at low
redshift, at fixed mass the relationship between (sub)halo overdensity
and mass overdensity is much more linear and close to unbiased for the
full set of subhalos than it is for the set of halos which just
formed. This effect is commonly understood as resulting from the fact
that the densest regions are usually fully contained in large mass
halos by low redshift, and thus cannot form new subhalos.

Consider now Figure \ref{biascomp}. The top panels show the rate of
formation $dG/dz$ of subhalos of masses $M= 10^{9} M_\odot$ on the
left and $10^{11} M_\odot$ on the right. The bottom panel shows as the
solid line the results for large-scale, linear bias of Equation
\ref{linbias}. The dotted line is the bias at formation of Equation
\ref{b_form}.  To demonstrate what is going on, I use the linear
continuity equation to extrapolate backwards from the $z=0$ results of
the solid line. This dashed line crosses the dotted line at around the
median of the $dG/dz$ distribution. The results for the bias at $z=0$
are similar what one would get by assuming that all the halos formed
at the median $dG/dz$ value.

This description {\it sounds} similar to the usual method suggested by
\citet{mo96a}, to {\it identify} halos at one redshift and evaluate
their clustering at another redshift. However, different mass halos
form at different times. Consider the top panel of Figure
\ref{massdep}, which shows as the solid lines the results of Equation
\ref{nlbias} as a function of mass for smoothing scales of $R=8$ and
30 $h^{-1}$ Mpc (as labeled), defining bias as $b =
\avg{\delta_{sh}\delta}/\avg{\delta\delta}$. For comparison, consider
the dashed lines, which show the bias as a function of mass using the
\citet{mo96a} results for halos identified at $z=1.5$ and observed at
$z=0$. In general, for my results the bias is less different than
unity, because the bias near the median time of formation is typically
closer to unity. Furthermore, the low mass halos which form very early
naturally debias according to the continuity equation.  The bias is
much less mass dependent under my assumptions than it is under the
assumption that all halos stop merging at some fixed redshift.

These results are extremely close to what one would get assuming all
subhalos were formed at the median of the $dG/dz$ distribution,
determine by setting
\begin{equation}
\delta_c^2(z_{\mathrm{med}}) = 2\sigma^2(M) \delta_{c,0}^2 \ln 2 +
\delta_c(z_{\mathrm{obs}}).
\end{equation}
The dashed lines in the bottom panel of Figure \ref{massdep} show the
mass dependence using the standard \citet{mo96a} treatment but setting
the redshift of identification at each mass to $z_{\mathrm{med}}$.

Finally, although at low mass the bias is a very weak function of mass
for the subhalos, recently assembled subhalos are in lower density
regions than older subhalos. Figure \ref{massage} plots the bias of
subhalos as a function of their mass and age, showing that at low
masses age is important while at high masses mass is important.
Accepting that galaxy luminosity is related to subhalo mass, and that
the optical color of a galaxy relates to the time of subhalo assembly,
this trend is qualitatively similar to the trend found in
\citet{hogg03b}. Those authors show that for low luminosity galaxies,
color is most closely related to galaxy overdensity, while for high
luminosity galaxies, luminosity is the most closely related to galaxy
overdensity.

However, the transformation from age of assembly and mass of a subhalo
to color and luminosity of a galaxy is by no means simple or even
necessarily one-to-one. In particular, the simplest interpretation of
the hierarchical merging scenario of galaxy formation is that the most
massive galaxies have assembled the most recently; however, it is
observationally the case that the most massive galaxies have the
oldest, reddest stellar populations. So if this scenario has any truth
at all, it must be the case that age of assembly and color are not
related at least for the most luminous galaxies, and so drawing firm
conclusions for the low mass subhalos would be premature. 

Other results of \citet{hogg03b} are more difficult to understand with
this picture, such as the large overdensities associated with red, low
luminosity galaxies. This discrepancy may indicate that such galaxies
preferentially survive in large mass clusters rather than in lower
mass groups.

\def\figsize{7.0cm} 
\def\figsiz{7.0cm}
\begin{figure} 
%\vskip\smtopskip
\centerline{\epsfxsize=\figsize\epsffile{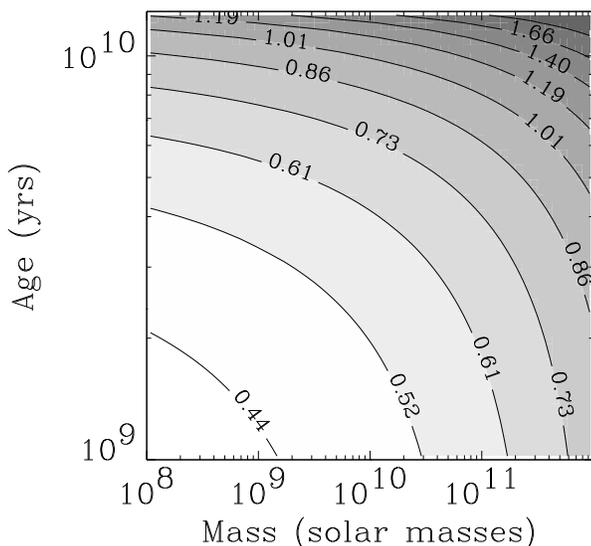}}
%\vskip\smbotskip
\caption[1]{\footnotesize% 
Contour plot of the bias of subhalos $b
=\avg{\delta_{sh}\delta}/\avg{\delta\delta}$ at redshift $z=0$ on 8
$h^{-1}$ Mpc smoothing scales as a function of mass and time of
assembly. At low masses time of assembly is more strongly related to
the large scale density field than it is at high masses.
\label{massage}}

\end{figure}

\section{Discussion}

An observational consequence of the model presented here is that the
clustering of galaxies should be a weak function of their mass,
especially at low masses.  Many papers have claimed that the average
clustering is a weak function of luminosity at luminosities less than
$L_\ast$ (\citealt{hamilton88a,giuricin01a,norberg02a, zehavi02a,
blanton03d,hogg03b}). Since dynamical and lensing studies suggest that
$L_\ast$ galaxies have total masses of order $10^{12} M_\odot$
(\citealt{mckay02a}), the observational results are in agreement with
Figure \ref{massdep}.  In addition, this model qualitatively explains
the trend in \citet{hogg03b} that at low luminosity, galaxy clustering
is a strong function of color.

These suggestive comparisons to observations inspire testing of the
assumptions used here with $N$-body simulations. Unfortunately,
identifying subhalos in simulations is difficult. While many
investigators are using high resolution simulations and identifying
subhalos, not much literature quotes directly subhalo bias with
respect to mass overdensity. The results of \citet{kravtsov99a} show
that the relationship between their subhalo population and the mass
overdensity is more linear and closer to $b=1$ than the \citet{mo96a}
prediction for halos, but that if they identify halos at
$z=z_{\mathrm{obs}}+1$ and observe them at $z_{\mathrm{obs}}$ the
formulae of \citet{mo96a} provide a reasonable fit. It would be
interesting use similar simulations to predict clustering as a
function of subhalo mass and to compare with the simple estimates I
present here.

Differences from this model in the observations or in the simulations
may indicate a breakdown of the assumption that subhalo survival
probability is independent of environment and may provide an insight
into the nature of that process.

A significant improvement of this model is to express the argument
here in terms of the ``halo model'' of galaxy formation predicting the
distribution of subhalos within each halo (see \citealt{berlind02a}
and references therein). This requires an extension of the typical
excursion set picture of Press-Schechter, and is implemented by Sheth
(in preparation).

\acknowledgments I thank Ravi Sheth for helping me along with this
project, David W. Hogg for useful discussions, and David H. Weinberg
for useful comments on the text. I am grateful for the hospitality of
the Department of Physics and Astronomy at the State University of New
York at Stony Brook, at which most of this work was done.

\end{document}